\documentclass[12pt,twoside, a4paper]{article}
\def\pd{\partial}
\def\mc{\mathcal}

\usepackage[dvips]{graphicx}
\usepackage{amssymb}
\usepackage{amssymb,amsmath}
\usepackage{graphicx}
\usepackage{dsfont}
\usepackage{caption}
\usepackage{subcaption}
\input{epsf.sty}  
\begin{document}
\begin{center}
\LARGE{\textbf{$dS_5$ vacua from matter-coupled 5D $N=4$ gauged supergravity}}
\end{center}
\vspace{1 cm}
\begin{center}
\large{\textbf{H. L. Dao}$^a$ and \textbf{Parinya Karndumri}$^b$}
\end{center}
\begin{center}
$^a$\textsl{Department of Physics, National University of Singapore,
3 Science Drive 2, Singapore 117551}\\
E-mail: hl.dao@u.nus.edu\\
$^b$\textsl{String Theory and Supergravity Group, Department
of Physics, Faculty of Science, Chulalongkorn University, 254 Phayathai Road, Pathumwan, Bangkok 10330, Thailand} \\
E-mail: parinya.ka@hotmail.com \vspace{1 cm}
\end{center}
\begin{abstract}
We study $dS_5$ vacua within matter-coupled $N=4$ gauged supergravity in five dimensions using the embedding tensor formalism. With a simple ansatz for solving the extremization and positivity of the scalar potential, we derive a set of conditions for the gauged supergravity to admit $dS_5$ as maximally symmetric background solutions. The results provide a new approach for finding $dS_5$ vacua in five-dimensional $N=4$ gauged supergravity and explain a number of notable features pointed out in previous works. These conditions also determine the form of the gauge groups to be $SO(1,1)\times G_{\textrm{nc}}$ with $G_{\textrm{nc}}$ being a non-abelian non-compact group. In general, $G_{\textrm{nc}}$ can be a product of $SO(1,2)$ and a smaller non-compact group $G'_{\textrm{nc}}$ together with (possibly) a compact group. The $SO(1,1)$ factor is gauged by one of the six graviphotons, that is singlet under $SO(5)\sim USp(4)$ R-symmetry. The compact parts of $SO(1,2)$ and $G'_{\textrm{nc}}$ are gauged by vector fields from the gravity and vector multiplets, respectively. In addition, we explicitly study $dS_5$ vacua for a number of gauge groups and compute scalar masses at the vacua. As in the four-dimensional $N=4$ gauged supergravity, all the $dS_5$ vacua identified here are unstable.      
\end{abstract}
\newpage
\tableofcontents
\newpage
\section{Introduction}
Finding de Sitter $(dS)$ vacua, with a positive cosmological constant, is one of the most interesting areas in gauged supergravities due to their importance in cosmology \cite{dS_cosmo1,dS_cosmo2,dS_cosmo3} and the dS/CFT correspondence \cite{dS_CFT}. Unlike the $AdS$ vacua which are rather common in gauged supergravities, $dS$ vacua are rare and occur only for a specific form of gauge groups. Moreover, embedding these vacua from lower-dimensional effective theories in string/M-theory is a highly non-trivial task, see for example \cite{dS_string1,dS_string2,dS_string3,dS_string4,flux_moduli,complete_Mink_vacua,dS_string6,dS6,dS_Sdual,dS8}, and a recent review \cite{Danielsson-18}.   
\\
\indent In four dimensions, de Sitter vacua are extensively studied due to their direct relevance for cosmology, see for example \cite{Hull-84,de-Roo-Panda-1,Trigiante,de-Roo-Panda-2,Roest-09}. In other dimensions, much less is known about this type of vacua. In this paper, we are interested in $dS_5$ vacua of $N=4$ gauged supergravity coupled to vector multiplets constructed in \cite{5D_N4_Dallagata,N4_gauged_SUGRA}. There are no $dS_5$ vacua in pure $N=4$ gauged supergravity of \cite{5D_N4_Romans}. $dS_5$ vacua of $N=4$ gauged supergravity coupled to two vector multiplets have already appeared in \cite{smet-PhD}. From a number of explicit examples, it has been pointed out that gauge groups admitting $dS_5$ vacua must contain non-compact abelian $SO(1,1)$ and non-compact non-abelian factors. To the best of our knowldege, these are the only known $dS_5$ vacua in the framework of matter-coupled $N=4$ gauged supergravity. There are also a number of $dS_5$ vacua in the maximal $N=8$ and the minimal $N=2$ gauged supergravities \cite{gunaydin-86,gunaydin-00,ogetbil-1,ogetbil-2,smet-05}. However, most of these $dS_5$ vacua are unstable except for a few examples in \cite{smet-05}. 
\\
\indent We will use the embedding tensor formalism to analyze the extremization and positivity of the scalar potential in terms of fermion-shift matrices. With a simple ansatz, we can derive a number of conditions the embedding tensor must satisfy in order for the scalar potential to admit a $dS_5$ vacuum. In general, solving these conditions subject to the qudratic constraint will determine the form of possible gauge groups. The analysis and the resulting conditions are very similar to the conditions for the existence of maximally supersymmetric $AdS_5$ vacua studied in \cite{AdS5_N4_Jan}. In a sense, our results can be regarded as a $dS_5$ analogue but are based on some assumption rather than supersymmetry. Although the conditions we impose are very restrictive, we will show that our results explicitly lead to the fact that the gauge groups must be of the form $SO(1,1)\times G_{\textrm{nc}}$ with $G_{\textrm{nc}}$ being a non-abelian and non-compact group as pointed out in \cite{smet-PhD}. In general, for $G_{\textrm{nc}}$ being a semisimple group, $G_{\textrm{nc}}$ can be a product of $SO(2,1)$ and a smaller non-compact group $G'_{\textrm{nc}}$.     
\\
\indent The paper is organized as follows. In section \ref{N4_SUGRA},
we review relevant formulae for computing the scalar potential of $N=4$ gauged supergravity in five dimensions coupled to vector multiplets in the embedding tensor formalism. In section \ref{dS5}, we derive the conditions that the embedding tensor needs to satisfy in order for the scalar potential to admit $dS_5$ vacua and determine a general form of gauge groups implied by these conditions. Some examples of these gauge groups are explicitly studied in section \ref{dS5_example}, and conclusions and comments are given in section \ref{conclusions}. We also include an appendix containing useful identities for $SO(5)$ gamma matrices and a collection of non-vanishing components of the embedding tensor for all gauge groups considered in this paper.  

\section{Five dimensional $N=4$ gauged supergravity coupled to vector multiplets}\label{N4_SUGRA} 
In this section, we review five dimensional $N=4$ gauged supergravity coupled to $n$ vector multiplets. We will only focus on formulae relevant for finding vacuum solutions. The detailed construction can be found in \cite{5D_N4_Dallagata} and \cite{N4_gauged_SUGRA}. 
\\
\indent The $N=4$ gravity multiplet consists of the graviton
$e^{\hat{\mu}}_\mu$, four gravitini $\psi_{\mu i}$, six vectors $A^0$ and
$A_\mu^m$, four spin-$\frac{1}{2}$ fields $\chi_i$ and one real
scalar $\Sigma$, the dilaton. Space-time and tangent space indices are denoted respectively by $\mu,\nu,\ldots =0,1,2,3,4$ and
$\hat{\mu},\hat{\nu},\ldots=0,1,2,3,4$. The $SO(5)\sim USp(4)$
R-symmetry indices are described by $m,n=1,\ldots, 5$ for the
$SO(5)$ vector representation and $i,j=1,2,3,4$ for the $SO(5)$
spinor or $USp(4)$ fundamental representation. The dilaton $\Sigma$ can be considered as a coordinate on $SO(1,1)\sim \mathbb{R}^+$ coset manifold. 
\\
\indent The vector multiplet contains a vector field $A_\mu$, four gaugini $\lambda_i$ and five scalars $\phi^m$. The $n$ vector
multiplets will be labeled by indices $a,b=1,\ldots, n$, and the component fields within these multiplets will be denoted by $(A^a_\mu,\lambda^{a}_i,\phi^{ma})$. From both gravity and vector multiplets, there are in total $6+n$ vector fields which will be collectively denoted by $A^{\mc{M}}_\mu=(A^0_\mu,A^M_\mu)=(A^0_\mu,A^m_\mu,A^a_\mu)$. The $5n$ scalar fields from the vector multiplets parametrize the $SO(5,n)/SO(5)\times SO(n)$ coset. To describe this coset manifold, we introduce a coset representative $\mc{V}_M^{\phantom{M}A}$ transforming under the global $SO(5,n)$ and the local $SO(5)\times SO(n)$ by left and right multiplications, respectively. We use indices $M,N,\ldots=1,2,\ldots , 5+n$ for global $SO(5,n)$ indices. The local $SO(5)\times SO(n)$ indices $A,B,\ldots$ will be split into $A=(m,a)$. We can accordingly write the coset representative as
\begin{equation}
\mc{V}_M^{\phantom{M}A}=(\mc{V}_M^{\phantom{M}m},\mc{V}_M^{\phantom{M}a}).
\end{equation}
The matrix $\mc{V}_M^{\phantom{M}A}$ is an element of $SO(5,n)$ and satisfies the relation
\begin{equation}
\eta_{MN}={\mc{V}_M}^A{\mc{V}_N}^B\eta_{AB}=-\mc{V}_M^{\phantom{M}m}\mc{V}_N^{\phantom{M}m}+\mc{V}_M^{\phantom{M}a}\mc{V}_N^{\phantom{M}a}
\end{equation}
with $\eta_{MN}=\textrm{diag}(-1,-1,-1,-1,-1,1,\ldots,1)$ being the $SO(5,n)$ invariant tensor. 
\\
\indent Gaugings are implemented by promoting a subgroup $G_0$ of the full global symmetry $G=SO(1,1)\times SO(5,n)$ to be a local symmetry. The most general gaugings can be described by using the embedding tensor. $N=4$ supersymmetry allows three components of the embedding tensor denoted by $\xi^{M}$, $\xi^{MN}=\xi^{[MN]}$ and $f_{MNP}=f_{[MNP]}$ \cite{N4_gauged_SUGRA}.  
The embedding tensor leads to minimal coupling of various fields via the covariant derivative
\begin{equation}
D_\mu=\nabla_\mu -A^M_\mu{f_M}^{NP}t_{NP}-A^0_\mu \xi^{NP}t_{NP}-A^M_\mu \xi^Nt_{MN}-A^M_\mu\xi_Mt_0
\end{equation}
in which $\nabla_\mu$ is the usual space-time covariant derivative and $t_{MN}=t_{[MN]}$ and $t_0$ are generators of $SO(5,n)$ and $SO(1,1)$, respectively. It should also be noted that $\xi^M$, $\xi^{MN}$ and $f_{MNP}$ include the gauge coupling constants, and $SO(5,n)$ indices $M,N,\ldots$ are lowered and raised by $\eta_{MN}$ and its inverse $\eta^{MN}$, respectively. 
\\
\indent In term of the embedding tensor, gauge generators ${X_{\mc{M}\mc{N}}}^{\mc{P}}={(X_{\mc{M}})_{\mc{N}}}^{\mc{P}}$ are given by
\begin{eqnarray}
{X_{MN}}^P&=&-{f_{MN}}^P-\frac{1}{2}\eta_{MN}\xi^P+\delta^P_{[M}\xi_{N]},\nonumber \\
{X_{M0}}^0&=&\xi_M,\qquad {X_{0M}}^N=-{\xi_M}^N\, .
\end{eqnarray}
To ensure that these generators form a closed subalgebra of $G$
\begin{equation}
\left[X_{\mc{M}},X_{\mc{N}}\right]=-{X_{\mc{M}\mc{N}}}^{\mc{P}}X_{\mc{P}},
\end{equation} 
the embedding tensor must satisfy the following quadratic constraint
\begin{eqnarray}
\xi^M\xi_M&=&0,\qquad \xi_{MN}\xi^N=0,\qquad f_{MNP}\xi^P=0,\nonumber\\
3f_{R[MN}{f_{PQ]}}^R&=&2f_{[MNP}\xi_{Q]},\qquad {\xi_M}^Qf_{QNP}=\xi_M\xi_{NP}-\xi_{[N}\xi_{P]M}\, .\label{QC}
\end{eqnarray}
\indent Since we are only interested in maximally symmetric solutions, we will set all fields but the metric and scalars to zero. In this case, the bosonic Lagrangian of a general gauged $N=4$ supergravity coupled to $n$ vector multiplets can be written as
\begin{equation}
e^{-1}\mc{L}=\frac{1}{2}R-\frac{3}{2}\Sigma^{-2}\pd_\mu \Sigma \pd^\mu \Sigma +\frac{1}{16} \pd_\mu M_{MN}\pd^\mu
M^{MN}-V
\end{equation}
where $e$ is the vielbein determinant. The scalar potential is given by
\begin{eqnarray}
V&=&-\frac{1}{4}\left[f_{MNP}f_{QRS}\Sigma^{-2}\left(\frac{1}{12}M^{MQ}M^{NR}M^{PS}-\frac{1}{4}M^{MQ}\eta^{NR}\eta^{PS}\right.\right.\nonumber \\
& &\left.+\frac{1}{6}\eta^{MQ}\eta^{NR}\eta^{PS}\right) +\frac{1}{4}\xi_{MN}\xi_{PQ}\Sigma^4(M^{MP}M^{NQ}-\eta^{MP}\eta^{NQ})\nonumber \\
& &\left.
+\frac{\sqrt{2}}{3}f_{MNP}\xi_{QR}\Sigma M^{MNPQRS}+\xi^M\xi^N\Sigma^{-2}M^{MN}\right]
\end{eqnarray}
with $M^{MN}$ being the inverse of a symmetric matrix $M_{MN}$ defined by
\begin{equation}
M_{MN}=\mc{V}_M^{\phantom{M}m}\mc{V}_N^{\phantom{M}m}+\mc{V}_M^{\phantom{M}a}\mc{V}_N^{\phantom{M}a}\, .
\end{equation}
$M^{MNPQRS}$ is obtained from raising indices of $M_{MNPQR}$ defined by
\begin{equation}
M_{MNPQR}=\epsilon_{mnpqr}\mc{V}_{M}^{\phantom{M}m}\mc{V}_{N}^{\phantom{M}n}
\mc{V}_{P}^{\phantom{M}p}\mc{V}_{Q}^{\phantom{M}q}\mc{V}_{R}^{\phantom{M}r}\, .
\end{equation}
\indent Equivalently, the scalar potential can also be written in terms of fermion-shift matrices as 
\begin{equation}
-\frac{1}{4}\Omega^{ij}V=A_1^{ik}{A_1^j}_k-A_2^{ik}{A_2^j}_k-A_2^{aik}{A_2^{aj}}_k
\end{equation}
or, after a contraction of $i$ and $j$ indices,
\begin{equation}
V=-A_1^{ij}A_{1ij}+A_2^{ij}A_{2ij}+A_2^{aij}A^a_{2ij}\, .
\end{equation}
The fermion shift matrices are in turn defined by
\begin{eqnarray}
A_1^{ij}&=&\frac{1}{\sqrt{6}}\left(-\zeta^{(ij)} + 2\rho^{(ij)}\right), \\
A_2^{ij}&=&\frac{1}{\sqrt{6}}\left(\zeta^{(ij)} + \rho^{(ij)}+\frac{3}{2}\tau^{[ij]} \right), \\
A_2^{aij}&=&\frac{1}{2}\left(-\zeta^{a[ij]} + \rho^{a(ij)}-\frac{\sqrt{2}}{4}\tau^a\Omega^{ij}\right)
\end{eqnarray}
where
\begin{eqnarray}
\tau^{[ij]}&=&\Sigma^{-1}{\mc{V}_M}^{ij}\xi^M,\qquad \tau^a=\Sigma^{-1}{\mc{V}_M}^a\xi^M,\qquad \zeta^{(ij)}=\sqrt{2}\Sigma^2\Omega_{kl}{\mc{V}_M}^{ik}{\mc{V}_N}^{jl}\xi^{MN},\nonumber  \\ \zeta^{a[ij]}&=&\Sigma^2{\mc{V}_M}^a{\mc{V}_N}^{ij}\xi^{MN},\qquad
\rho^{(ij)} =-\frac{2}{3}\Sigma^{-1}{\mc{V}^{ik}}_M{\mc{V}^{jl}}_N{\mc{V}^P}_{kl}{f^{MN}}_P, \nonumber \\
\rho^{a(ij)}&=&\sqrt{2}\Sigma^{-1}\Omega_{kl}{\mc{V}_M}^a{\mc{V}_N}^{ik}{\mc{V}_P}^{jl}f^{MNP}\, .\label{definition}
\end{eqnarray}
We have explicitly shown the symmetry of each tensor for later convenience. Note also that lowering and raising of $USp(4)$ indices $i,j,\ldots$ with the symplectic form $\Omega_{ij}$ and its inverse $\Omega^{ij}$ correspond to complex conjugate for example $A_{1ij}=\Omega_{ik}\Omega_{jl}A_1^{kl}=(A_1^{ij})^*$. 
\\
\indent ${\mc{V}_M}^{ij}$ is related to ${\mc{V}_M}^m$ by $SO(5)$ gamma matrices ${\Gamma_{mi}}^j$. As in \cite{AdS5_N4_Jan}, we define $\mc{V}_M^{\phantom{M}ij}$ by
\begin{equation}
{\mc{V}_M}^{ij}={\mc{V}_M}^{m}\Gamma^{ij}_m
\end{equation}
where $\Gamma^{ij}_m=\Omega^{ik}{\Gamma_{mk}}^j$. ${\mc{V}_M}^{ij}$ satisfies the following relations
\begin{equation}
{\mc{V}_M}^{ij}=-{\mc{V}_M}^{ji}\qquad \textrm{and} \qquad {\mc{V}_M}^{ij}\Omega_{ij}=0\, .
\end{equation}
Similarly, the inverse element ${\mc{V}_{ij}}^M$ can be written as
\begin{equation}
{\mc{V}_{ij}}^M={\mc{V}_m}^M(\Gamma^{ij}_m)^*={\mc{V}_m}^M\Gamma_{m}^{kl}\Omega_{ki}\Omega_{lj}\, .
\end{equation}

\section{de Sitter vacua of $N=4$ five-dimensional gauged supergravity}\label{dS5}
We now consider gauge groups that lead to de Sitter vacua. As in \cite{AdS5_N4_Jan}, it is useful to introduce ``dressed'' components of the embedding tensor
\begin{eqnarray}
\xi^A&=&\langle{\mc{V}_M}^A\rangle \xi^M,\qquad \xi^{AB}  =\langle {\mc V_M}^A\rangle \langle {V_N}^B\rangle \xi^{MN}, 
\nonumber \\
f^{ABC} &=& \langle {\mc V_M}^A\rangle \langle {\mc V_N}^B\rangle\langle {\mc V_P}^C\rangle f^{MNP}
\end{eqnarray}
where, as in \cite{AdS5_N4_Jan}, $\langle \,\,\, \rangle$ means the quantity inside is evaluated at the vacuum. Using the splitting of the index $A=(m,a)$, we have the following components of the embedding tensor under the decomposition $SO(5,n)\rightarrow SO(5)\times SO(n)$
\begin{equation}
\xi^A=(\xi^m,\xi^a),\quad \xi^{AB} = (\xi^{mn}, \xi^{ma}, \xi^{ab}),\quad f^{ABC}=(f^{mnp}, f^{mna}, f^{abm}, f^{abc}).
\end{equation}
In subsequent snalysis, we will determine the form of gauge groups that lead to $dS_5$ vacua. With some assumption, the  analysis is very similar to that of the supersymmetric $AdS_5$ vacua studied in \cite{AdS5_N4_Jan}.  
\\
\indent In order to have $dS_5$ vacua, we require that
\begin{equation}
\langle \delta V\rangle=0\qquad \textrm{and}\qquad \langle V\rangle>0\, .\label{general_dS5_con}
\end{equation}
In terms of the fermion-shift matrices, these conditions read
\begin{eqnarray}
\langle \delta V \rangle&=&-2\langle \delta A^{ij}_1A_{1ij}\rangle+2\langle \delta A_2^{ij} A_{2ij}\rangle+2\langle\delta A^{aij}_2 A^a_{2ij} \rangle =0,\\
\textrm{and}& & \qquad \langle A^{ij}_2A_{2ij}\rangle+\langle A^{aij}_2A^a_{2ij}\rangle>\langle A^{ij}_1A_{1ij}\rangle\, .
\end{eqnarray}
There are various ways to satisfy these relations. To make the analysis tractable, we will restrict ourselves to the following two possibilities:
\begin{enumerate}
\item $\langle A_1^{ij} \rangle=0$, $\langle A^{aij}_2 \rangle=0$ and $\langle A^{ik}_2 A_{2jk} \rangle=\frac{1}{4}|\mu|^2\delta^i_j$ with $A_{2ij}\delta A^{ij}_2=0$.
\item $\langle A_1^{ij} \rangle=0$, $\langle A^{ij}_2 \rangle=0$ and $\langle A^{aik}_2 A^a_{2jk} \rangle=\frac{1}{4}|\mu|^2\delta^i_j$ with $A^a_{2ij}\delta A^{aij}_2=0$.
\end{enumerate} 
$|\mu|^2$ denotes the cosmological constant, the value of the scalar potential at the vacuum $V_0$. 

\subsection{$\langle A_1^{ij} \rangle=0$, $\langle A^{aij}_2 \rangle=0$ and $\langle A^{ik}_2 A_{2jk} \rangle=\frac{1}{4}|\mu|^2\delta^i_j$}
We begin with the first possibility. Since $A^{aij}_2$ consists of three representations of $USp(4)$ namely $\mathbf{10}$, $\mathbf{5}$ and $\mathbf{1}$ corresponding to $\rho^{a(ij)}$, $\zeta^{a[ij]}$ and $\Omega^{ij}\tau^a$, respectively, the condition $\langle A^{aij}_2\rangle=0$ then implies that these components must vanish separately
\begin{equation}
\langle \rho^{a(ij)} \rangle=\langle \zeta^{a[ij]} \rangle=\langle \tau^a\rangle=0\, .\label{con_1}
\end{equation}  
The condition $\langle A^{ij}_1\rangle=0$ gives
\begin{equation}
\langle \zeta^{(ij)}\rangle=2\langle \rho^{(ij)}\rangle\, .\label{zeta_rho}
\end{equation}
Using the definitions in \eqref{definition}, we have
\begin{eqnarray}
\tau^a&=&\Sigma^{-1}\xi^a,\qquad \zeta^{a[ij]}=\Sigma^2\xi^{am}\Gamma^{ij}_m,\nonumber \\
\rho^{a(ij)}&=&\sqrt{2}\Sigma^{-1}\Omega_{kl}f^{amn}\Gamma_m^{ik}\Gamma^{jl}_n=-\sqrt{2}\Sigma^{-1}f^{amn}(\Gamma_{mn})^{ij}\, .\label{zeta_rho_def}
\end{eqnarray}
The conditions in \eqref{con_1} then imply that
\begin{equation}
\xi^a=\xi^{am}=f^{amn}=0
\end{equation}
due to the non-vanishing $\langle \Sigma\rangle$ and $\Gamma^{ij}_m$ and $\Gamma_{mn}^{ij}$ being all linearly independent. Using the first condition of the quadratic constraint \eqref{QC}, we find that
\begin{equation}
\xi^M\xi_M=-\xi^m\xi_m+\xi^a\xi_a=\xi^m\xi_m=0
\end{equation}
which implies $\xi^m=0$. Accordingly, we are left with $\xi^M=0$.
\\
\indent With $\xi^M=0$ together with \eqref{zeta_rho}, we have $\tau^{[ij]}=\Sigma^{-1}\xi^{m}\Gamma^{ij}_m=0$ and 
\begin{equation}
A_2^{ij}=\frac{3}{2\sqrt{6}}\zeta^{(ij)}=\frac{\sqrt{3}}{2}\Sigma^2\xi^{mn}(\Gamma_{mn})^{ij}\, .
\end{equation}
Using the identity given in \eqref{eq:a7} and the fact that $(A^{ij}_2)^*=A_2^{kl}\Omega_{ki}\Omega_{lj}$, we find, from $A_2^{ik}A_{2jk}=\frac{1}{4}|\mu|^2 \delta^i_j$,
\begin{eqnarray}
\frac{1}{4}|\mu|^2\delta^i_j&=&-\frac{3}{2}\langle \Sigma^4\rangle \xi^{mn}\xi^{pq}{(\Gamma_{mn}\Gamma_{pq})^i}_j=-\frac{3}{4}\langle \Sigma^4\rangle \xi^{mn}\xi^{pq}{\{\Gamma_{mn},\Gamma_{pq}\}^i}_j\nonumber \\
&=&-\frac{3}{2}\langle \Sigma^4\rangle \xi^{mn}\xi^{pq}{(\Gamma_{mnpq})^i}_j+3\langle \Sigma^4\rangle \xi^{mn}\xi_{mn}\delta^i_j
\end{eqnarray}
which gives
\begin{equation}
\xi^{[mn}\xi^{pq]}=0\qquad \textrm{and}\qquad \langle \Sigma^4\rangle\xi^{mn}\xi_{mn}=\frac{|\mu|^2}{12}\, .
\end{equation}
In addition, the condition \eqref{zeta_rho} gives
\begin{equation}
3\sqrt{2}\xi_{qr}\langle \Sigma^3\rangle=-2\epsilon_{mnpqr}f^{mnp}\, .
\end{equation}
We can easily see that, apart from numerical differences, these conditions have a very similar structure to those for the existence of supersymmetric $AdS_5$ vacua. However, we still need to check whether these conditions extremize the potential. Note that for the $AdS_5$ case, the potential is automatically extremized since in this case we have 
\begin{equation}
\delta A_1^{ij}=-2\Sigma^{-1}A^{ij}_2\delta\Sigma+\frac{2}{\sqrt{3}}\Omega_{kl}\Gamma_m^{k(i}A^{j)l}_{2a}\delta \phi^{ma}
\end{equation}
where we have introduced the variations of the coset representative ${\mc{V}_M}^A$ with respect to scalars $\phi^{ma}$
\begin{equation}
\delta {\mc{V}_M}^m={\mc{V}_M}^a\delta \phi^{ma}\qquad \textrm{and}\qquad \delta {\mc{V}_M}^a={\mc{V}_M}^m\delta \phi^{ma}\, .
\end{equation}
With $\langle A^{ij}_2\rangle=\langle A^{aij}_2\rangle=0$, we immediately see that $\langle \delta V\rangle =-2\langle A_{1ij}\delta A_1^{ij}\rangle=0$.
\\
\indent However, in the present case, with $\langle A^{ij}_1\rangle=\langle A^{aij}_2\rangle=0$, we find 
\begin{equation}
\langle \delta_\Sigma V\rangle=2\langle A_{2ij}\delta_\Sigma A_2^{ij}\rangle=-3\langle \Sigma^{-1}\rangle\langle \rho^{(ij)}\rho_{(ij)}\rangle\delta \Sigma
\end{equation}
where we have used the relation
\begin{equation}
\delta_\Sigma \rho^{(ij)}=-\Sigma^{-1}\rho^{(ij)}\delta\Sigma\, .
\end{equation}    
We see that $\langle \delta V\rangle=0$ implies $\langle \rho^{(ij)}\rangle=\langle A^{ij}_2\rangle=0$. The cosmological constant then vanishes, and there are no possible $dS_5$ vacua. 
\\
\indent It is now useful to note that if all the above conditions extremized the potential, the structure of the resulting gauge groups would be the same as in the $AdS_5$ case namely $U(1)\times H$ with $H$ containing an $SU(2)$ subgroup gauged by three of the graviphotons. Therefore, the same gauge group would give two types of vacua, $AdS_5$ and $dS_5$, with different ratios of the gauge coupling constants for the $U(1)$ and $H$ factors. According to the explicit form of the scalar potential studied in \cite{5D_N4_flows} and \cite{5D_N4_BH}, this is not the case. There is no $dS_5$ vacuum for gauge groups of the form $U(1)\times H$ for any values of the coupling constants precisely in agreement with the above result.   

\subsection{$\langle A_1^{ij} \rangle=0$, $\langle A^{ij}_2 \rangle=0$ and $\langle A^{aik}_2 A^a_{2jk} \rangle=\frac{1}{4}|\mu|^2\delta^i_j$}
We now consider the second possibility with $\langle A_1^{ij} \rangle=\langle A^{ij}_2 \rangle=0$ and $\langle A^{aik}_2 A^a_{2jk} \rangle=\frac{1}{4}|\mu|^2\delta^i_j$. $A^{ij}_2$ consists of a symmetric and an anti-symmetric part. To satisfy the condition $\langle A^{ij}_2\rangle=0$, these two parts must separately vanish. This implies $\tau^{[ij]}=0$ or $\xi^m=0$, and, as in the previous case, the quadratic constraint $\xi^M\xi_M=0$ gives $\xi^a=0$. We again find that $\xi^M=0$. 
\\
\indent With $\xi^M=0$, the conditions $\langle A_1^{ij} \rangle=\langle A^{ij}_2 \rangle=0$ give
\begin{equation}
\langle \zeta^{ij}\rangle=\langle \rho^{ij}\rangle =0
\end{equation} 
which implies
\begin{equation}
\xi^{mn}=0\qquad \textrm{and}\qquad f^{mnp}=0\, .
\end{equation}
Using \eqref{zeta_rho_def} and the identity in \eqref{eq:a7} together with 
\begin{eqnarray}
\Gamma_m\Gamma_{pq}&=&\frac{1}{2}\{\Gamma_m,\Gamma_{pq}\}+\frac{1}{2}[\Gamma_m,\Gamma_{pq}]\nonumber\\
&=&\Gamma_{mpq}+\frac{1}{2}(\delta_{mp}\Gamma_q-\delta_{mq}\Gamma_p),
\end{eqnarray} 
we find, after some manipulation, that the condition $\langle A^{aik}_2 A^a_{2jk} \rangle=\frac{1}{4}|\mu|^2\delta^i_j$ gives
\begin{eqnarray}
|\mu|^2\delta^i_j&=&\left[\langle\Sigma^4\rangle\xi^{am}\xi_{am}+4\langle\Sigma^{-2}\rangle f^{amn}f_{amn}\right]\delta^i_j-2\langle \Sigma^{-2}f^{amn}f^{apq}\rangle {(\Gamma_{mnpq})^i}_j\nonumber \\
& &+2\sqrt{2}\langle\Sigma\rangle \xi^{am}f^{apq}{(\Gamma_{mpq})^i}_j+2\sqrt{2}\langle \Sigma\rangle \xi^{am}f^{amn}{(\Gamma_n)^i}_j\, .
\end{eqnarray}
After using \eqref{eq:a6}, we arrive at
\begin{eqnarray}
\langle \Sigma^4\rangle\xi^{am}\xi_{am}+4\langle\Sigma^{-2}\rangle f^{amn}f_{amn}&=&|\mu|^2,\label{A22_V0_1}\\
\xi^{a[m}f^{pq]a}&=&0\label{A22_V0_2}\\
\sqrt{2}\langle \Sigma^3\rangle\xi^{am}f_{amr}&=&f^{amn}f^{apq}\epsilon_{mnpqr}\, .\label{A22_V0_3}
\end{eqnarray}
\indent We then move to the extremization of the potential, $\langle \delta V\rangle=2\langle A^a_{2ij}\delta A^{aij}_2\rangle=0$, which gives
\begin{equation}
\langle\rho^{a(ij)}\delta_\Sigma\rho^a_{(ij)}\rangle+\langle\zeta^{a[ij]}\delta_\Sigma \zeta^a_{[ij]}\rangle=0
\end{equation}
and
\begin{equation}
\langle\rho^{a(ij)}\delta_\phi\rho^a_{(ij)}\rangle+\langle\zeta^{a[ij]}\delta_\phi \zeta^a_{[ij]}\rangle=0\, .\label{vary_con1}
\end{equation}
With the relations
\begin{equation}
\delta_\Sigma \zeta^{a[ij]}=2\Sigma^{-1}\zeta^{a[ij]}\delta \Sigma\qquad \textrm{and}\qquad \delta_\Sigma\rho^{a(ij)}=-\Sigma^{-1}\rho^{a(ij)}\delta\Sigma,
\end{equation}
the first condition gives
\begin{equation}
\langle\rho^{a(ij)}\rho^a_{(ij)}\rangle=2\langle\zeta^{a[ij]}\zeta^a_{[ij]}\rangle\, .\label{rho_a_zeta_a}
\end{equation}
From the definitions \eqref{definition}, we can easily derive the following relations
\begin{eqnarray}
\delta_\phi \zeta^{a[ij]}&=&\Sigma^2\xi^{mn}\Gamma_n^{ij}\delta\phi^{ma}+\Sigma^2 \Gamma^{ij}_m\xi^{ab}\delta\phi^{mb},\\
\delta_\phi \rho^{a(ij)}&=&\sqrt{2}\Sigma^{-1}f^{mnp}\Omega_{kl}\Gamma_m^{ik}\Gamma_n^{jl}\delta\phi^{pa}-2\sqrt{2}\Sigma^{-1}f^{abn}
\Omega_{kl}\Gamma^{k(i}_m\Gamma^{j)l}_n\delta\phi^{mb}\, .
\end{eqnarray}
After some manipulation together with \eqref{eq:a2}, we find that the condition \eqref{vary_con1} gives
\begin{equation}
\langle \Sigma^6\rangle\xi^{am}\xi^{ab}=4f^{amn}f^{mab}\label{xi_ab_f_mab}
\end{equation}
where we have used the previous results $\xi^{mn}=0$ and $f^{mnp}=0$.
\\
\indent We now need to find solutions to all of these conditions subject to the quadratic constraints \eqref{QC}. With $\xi^M=0$, the quadratic constraint simplifies considerably 
\begin{equation}
f_{R[MN}{f_{PQ]}}^R=0\qquad \textrm{and}\qquad {\xi_M}^Qf_{QNP}=0\, .\label{quadratic_constraint_sim}
\end{equation}
By using $SO(5,n)$ generators of the form ${(t_{MN})_P}^Q=\delta^Q_{[M}\eta_{N]P}$, we find that the gauge generators in the fundamental representation of $SO(5,n)$ are given by 
\begin{equation}
{(X_M)_N}^P=-{f_M}^{QR}{(t_{QR})_N}^P={f_{MN}}^P\quad \textrm{and}\quad {(X_0)_N}^P=-\xi^{QR}{(t_{QR})_N}^P={\xi_N}^P\label{gauge_gennerator_vector}
\end{equation}                     
where we have set $\xi^M=0$. Note also that, for $\xi^{M}=0$, the gauge group is entirely embedded in $SO(5,n)$. The quadratic constraint then implies that $[X_0,X_M]=0$, so $X_0$ generates an abelian factor. On the other hand, $X_M$ in general generate a non-abelian group with structure constants $f_{MNP}$, and the first quadratic constraint in \eqref{quadratic_constraint_sim} gives the corresponding Jacobi's identity.
\\
\indent Using the relations $\xi^{mn}=f^{mnp}=0$, the second relation in \eqref{quadratic_constraint_sim} gives
\begin{eqnarray}
{\xi_m}^af_{anp}={\xi_m}^af_{abc}={\xi_m}^af_{abn}={\xi_a}^bf_{bmn}=0,\label{Jaco_1}\\
{\xi_a}^bf_{bcd}+{\xi_a}^mf_{mbc}=0,\\
{\xi_a}^bf_{bcm}+{\xi_a}^nf_{mnc}=0\, .
\end{eqnarray}
We now see that the first condition in \eqref{Jaco_1} automatically solves \eqref{A22_V0_2}. Using components $(mnpq)$ of the first quadratic constraint in \eqref{quadratic_constraint_sim}, we find from \eqref{A22_V0_3} that
\begin{equation}  
\xi^{am}f_{amn}=0\, .\label{xi_f}
\end{equation}
\indent We now study consequences of all the above conditions on the structure of gauge groups. First of all, using the relation \eqref{rho_a_zeta_a} in \eqref{A22_V0_1}, we find that 
\begin{equation}
\langle \Sigma^4\rangle \xi^{am}\xi_{am}=2\langle\Sigma^{-2}\rangle f^{amn}f_{amn}=\frac{1}{3}|\mu|^2\, .
\end{equation}
This result implies that, for $|\mu|^2\neq 0$, we must have $\xi^{am}\neq 0$ and $f^{amn}\neq 0$. Therefore, the abelian generator $X_0$ corresponds to a non-compact group $SO(1,1)$, and the non-abelian part must be necessarily non-compact. This is in agreement with the result of \cite{smet-PhD} in which it has been pointed out, based on a few explicit examples, that only gauge groups of the form $SO(1,1)\times G_{\textrm{nc}}$ lead to $dS_5$ vacua. With a simple assumption, we are able to derive this result. 
\\
\indent From equation \eqref{xi_f}, we see that $\xi^{am}$ and $f^{amn}$ must be non-vanishing for different values of indices $m$ and $a$. This also implies that gauge groups of the form $SO(1,1)\times G_{\textrm{nc}}$ are only possible for $n\geq 2$ (at least two different values of $a$) as shown by explicit computation in \cite{smet-PhD} that the existence of $dS_5$ vacua requires at least two vector multiplets. We now determine possible gauge groups allowed by the remaining conditions. 
\\
\indent We begin with the simplest possibility namely $\xi^{ab}=f_{mab}=f_{abc}=0$. These components are not constrained by the existence of $dS_5$ vacua, so we can have $dS_5$ vacua regardless of their values. Since in this case we have $f_{abc}=0$, the compact part of $G_{\textrm{nc}}$ must be an abelian $SO(2)$ group. We find that $f_{amn}$ must correspond to $SO(2,1)$ group. The full gauge group then takes the form of a product $SO(1,1)\times SO(2,1)$.
\\
\indent For $\xi^{ab}=f_{abc}=0$ but $f_{mab}\neq 0$, the quadratic constraint implies that $f_{mab}$ and $f_{amn}$ cannot have common indices. Therefore, these components generate two separate non-compact groups with $f_{amn}$ corresponding to $SO(2,1)$ as in the previous case. Since the existence of $dS_5$ vacua requires $f_{mnp}=0$, the compact subgroup of the non-compact group corresponding to $f_{mab}$ must be $SO(2)$, hence $f_{mab}$ generate another $SO(2,1)'$ factor. However, it should be noted that the compact parts of the two $SO(2,1)$'s are embedded in the matter and R-symmetry directions, respectively. The full gauge group in this case is then given by $SO(1,1)\times SO(2,1)\times SO(2,1)'$.
\\
\indent For $\xi^{ab}=f_{mab}=0$ but $f_{abc}\neq 0$, we have the following non-vanishing components of the embedding tensors
\begin{equation}
\xi^{\tilde{a}\tilde{m}},\qquad f_{a'm'n'},\qquad f_{a'b'c'}\label{example_case}
\end{equation}
in which we have split indices as $m=(\tilde{m},m')$ and $a=(\tilde{a},a')$. In this case, $f_{a'b'c'}$ together with $f_{a'm'n'}$ form a non-compact group $\tilde{G}_{\textrm{nc}}$ whose compact subgroup $\tilde{H}_\textrm{c}$ is generated by $f_{a'b'c'}$. The quadratic constraint gives the standard Jacobi's identity for $\tilde{H}_\textrm{c}$. In general, there can be a subspace in which $f_{a'b'c'}$ do not have common indices with $f_{a'm'n'}$. We will denote these components as $f_{a''b''c''}$ with $f_{a''m'n'}=0$. In this case, $f_{a''b''c''}$ form a separate compact group $H_{\textrm{c}}$ while $f_{a'm'n'}$ and $f_{a'b'c'}$ still generate a smaller non-compact group $G_{\textrm{nc}}$. The full gauge group can be written as $SO(1,1)\times G_{\textrm{nc}}\times H_{\textrm{c}}$. 
\\
\indent If in addition $f_{m'a'b'}\neq 0$, we have an extra factor of $SO(2,1)'$, so a general gauge group takes the form of
\begin{equation}
SO(1,1)\times SO(2,1)'\times G_{\textrm{nc}}\times H_{\textrm{c}}
\end{equation} 
with the compact parts of $SO(2,1)$ and $G_{\textrm{nc}}$ are embedded along the R-symmetry and matter directions, respectively. 
\\
\indent The non-vanishing $\xi^{ab}$ on the other hand does not change the structure of the gauge groups. 

\section{$dS_5$ vacua from different gauge groups}\label{dS5_example}
In this section, we will study $N=4$ gauged supergravity with gauge groups that lead to $dS_5$ vacua identified in the previous section. We will restrict ourselves to a simple case of $\xi^{ab}=f^{mab}=0$ given in \eqref{example_case} and consider only semisimple gauge groups. In general, the $SO(1,1)$ factor can be embedded as a diagonal subgroup $SO(1,1)^{(d)}_{\textrm{diag}}\sim(SO(1,1)^1\times \ldots \times SO(1,1)^d)_{\textrm{diag}}$ in which each $SO(1,1)$ factor is characterized by the values of indices $\tilde{m}$ and $\tilde{a}$. 
\\
\indent In order to have non-vanishing $f_{m'n'a'}$, we must have $d\leq 3$ to allow for the antisymmetry in $m'$ and $n'$ indices since $1\leq m',n'\leq 5-d$. An explicit form of the gauge generators is given by
\begin{eqnarray}
{(X_0)_M}^N&=&\left(\begin{array}{c|c|c|c}
    0 & 0 & {\xi_{\tilde{m}}}^{\tilde{b}} & 0 \\ \hline
    0 & 0 & 0 & 0 \\ \hline
    {\xi_{\tilde{a}}}^{\tilde{n}} & 0 & 0 & 0 \\ \hline
     0 & 0 & 0 & 0 
\end{array} \right),\quad \tilde{m}=1,\ldots, d;\, \tilde{a}=1,\ldots, \tilde{d}, \nonumber \\
{(X_{m'})_N}^P&=&\left(\begin{array}{c|c|c|c}
    0 & 0 & 0 & 0 \\ \hline
    0 & 0 & 0 & {f_{m'n'}}^{b'} \\ \hline
    0 & 0 & 0 & 0 \\ \hline
     0 & {f_{m'a'}}^{p'} & 0 & 0 
\end{array} \right),\quad m'=1,\ldots, 5-d;\, a'=1,\ldots, n-\tilde{d}, \nonumber \\
{(X_{a'})_N}^P&=&\left(\begin{array}{c|c|c|c}
    0 & 0 & 0 & 0 \\ \hline
    0 & {f_{a'm'}}^{n'} & 0 & 0 \\ \hline
    0 & 0 & 0 & 0 \\ \hline
     0 & 0 & 0 & {f_{a'b'}}^{c'} 
\end{array} \right) .
\end{eqnarray}
Generators $X_{m'}$ and $X_{a'}$ correspond to non-compact and compact generators of $G_{\textrm{nc}}$, respectively. Note also that we can have at most $5-d$ non-compact generators for $G_{\textrm{nc}}$. We now discuss possible gauge groups for different values of $d$. 
\\
\indent For $d=1$, $G_{\textrm{nc}}$ is a non-compact subgroup of $SO(4,n-1)$ with at most four non-compact generators. We can gauge the following groups
\begin{eqnarray}\label{eq:eligible-2}
SO(1,1)\times \begin{cases} SO(2,1), & n\geq 2 \\ SO(3,1), & n\geq 4\\
SO(2,1)^2\sim SO(2,2), & n\geq 3 \\  SU(2,1), &  n\geq 5\\ SO(4,1),  & n\geq 7\end{cases}\, .
\end{eqnarray}
\indent For $d=2$ and $d=3$, we have the following admissible gauge groups
\begin{eqnarray}
& &SO(1,1)^{(2)}_\text{diag}\times SO(2,1),\qquad n\geq 3, \nonumber \\ 
& &SO(1,1)^{(2)}_\text{diag}\times SO(3,1),\qquad n\geq 5, \nonumber \\ 
& &SO(1,1)^{(3)}_\text{diag}\times SO(2,1) ,\qquad n\geq 4\, .
\end{eqnarray}
We will explicitly study scalar potentials arising from these gauge groups and their $dS_5$ vacua. In most of the following analysis, we mainly work in the case of $n=5$ vector multiplets for definiteness with an exception of $SO(1,1)\times SO(4,1)$ gauge group that requires $n=7$. In addition, in many cases, we will consider only non-vanishing dilaton due to the complexity of including scalars from vector multiplets. In all of these gauge groups we have explicitly checked that the conditions $\langle A^{ij}_1\rangle=\langle A^{ij}_2\rangle=0$ and $\langle A^{aij}_2A^a_{2kj}\rangle=\frac{1}{4}V_0\delta^i_k$ are satisfied. We note here that $SO(1,1)\times SO(2,1)$ and $SO(1,1)^{(2)}_{\textrm{diag}}\times SO(2,1)$ gauge groups have already been studied in \cite{smet-PhD}. However, all the remaining gauge groups are new and arise from our analysis given in the previous section.
\\
\indent To compute the scalar potential, we need an explicit form of the coset representative for $SO(5,n)/SO(5)\times SO(n)$. It is convenient to define a basis of $GL(5+n,\mathbb{R})$ matrices
\begin{equation} 
(e_{MN})_{PQ}=\delta_{MP}\delta_{NQ}
\end{equation}
in terms of which $SO(5,n)$ non-compact generators are given by
\begin{equation}
Y_{ma}=e_{m,a+5}+e_{a+5,m},\qquad m=1,2,\ldots, 5,\qquad a=1,2,\ldots, n\, .
\end{equation}  
To adopt the normalization used in \cite{N4_gauged_SUGRA} with
\begin{equation}
\eta_{MN}=-{\mc{V}_M}^{ij}\mc{V}_{Nij}+{\mc{V}_M}^a{\mc{V}_N}^a,
\end{equation}
we will use a slightly different definition of ${\mc{V}_M}^{ij}$ and ${\mc{V}_{ij}}^M$ 
\begin{equation}
{\mc{V}_M}^{ij}=\frac{1}{2}{\mc{V}_M}^{m}\Gamma^{ij}_m\qquad \textrm{and}\qquad {\mc{V}_{ij}}^M=\frac{1}{2}{\mc{V}_m}^M(\Gamma^{ij}_m)^*\, .
\end{equation}
The explicit representation of the $SO(5)$ gamma matrices is chosen to be
\begin{eqnarray}
\Gamma_1&=&-\sigma_2\otimes \sigma_2,\qquad \Gamma_2=i\mathbb{I}_2\otimes \sigma_1,\qquad \Gamma_3=\mathbb{I}_2\otimes \sigma_3,\nonumber\\
\Gamma_4&=&\sigma_1\otimes \sigma_2,\qquad \Gamma_5=\sigma_3\otimes \sigma_2
\end{eqnarray}
where $\sigma_i$, $i=1,2,3$, are the usual Pauli matrices. 

\subsection{$SO(1,1)\times SO(2,1)$ gauge group}
We begin with the simplest possible gauge group $SO(1,1)\times SO(2,1)$. The embedding of $SO(1,1)$ factor is given by
\begin{equation}
\xi^{MN}=g_1(\delta^5_M\delta^6_N-\delta^5_M\delta^6_N)\label{SO1_1_embedding}
\end{equation}  
while the $f_{MNP}$ describing the embedding of $SO(2,1)$ has the following independent non-vanishing component  
\begin{eqnarray}
f_{238} = -g_2 \, .
\end{eqnarray}
This means that the compact part $SO(2)\subset SO(2,1)$ is generated by $X_8$ while the two noncompact generators are embedded in the R-symmetry directions $M=2,3$. Note also that this embedding tensor manifestly satisfies the quadratic constraint \eqref{quadratic_constraint_sim}.
\\
\indent At the vacuum, $SO(1,1)\times SO(2,1)$ gauge symmetry will be broken to its compact subgroup $SO(2)$. In principle, we can study the scalar potential for scalars which are singlets under this residual $SO(2)$ symmetry. However, the number of these singlet scalars is rather large, $15$ singlets in this case. Therefore, we will mainly analyze the scalar potential only to second-order in fluctuations of scalar fields and give the explicit form of the potential only for the dilaton $\Sigma$ non-vanishing. With all scalars from vector multiplets set to zero, the scalar potential is given by
\begin{eqnarray}
V=\frac{2g_2^2 + g_1^2 \Sigma^6}{4\Sigma^2}\, .
\end{eqnarray}
This potential admits a unique critical point at 
\begin{eqnarray}
\Sigma = \left(\frac{g_2}{g_1}\right)^{1/3},\qquad V_0 = \frac{3}{4}(g_1 g_2^2)^{2/3}\, .
\end{eqnarray}
\indent The value of the dilaton $\Sigma$ at the vacuum can be rescaled to $\Sigma=1$ by setting $g_2= g_1$ which gives the $dS_5$ vacuum in a simple form
\begin{equation}
\Sigma=1, \qquad V_0 =\frac{3}{4}g^2_1\, .
\end{equation}
The linearized analysis of the full scalar potential for all $26$ scalars shows that this is also the critical point of the full potential with scalar masses given by
\begin{equation}
\begin{array}{ccccccccc}
m^2 L^2= &16_{\times 1}, & 2(1+\sqrt{33})_{\times 2}, & 12_{\times 2}, & 2(1-\sqrt{33})_{\times 2}, &8_{\times 6}, &4_{\times 4}, & 0_{\times 9} \, .
\end{array}
\end{equation}
The number in the subscripts of each mass value is the multiplicity of that mass value at the critical point, and the $dS_5$ radius is given by $L^2 = 6/V_0$. This $dS_5$ critical point is unstable because of the negative mass value $2(1-\sqrt{33})$. The mass value $16$ corresponds to that of $\Sigma$, and the nine massless scalars contain three Goldstone bosons of the symmetry breaking $SO(1,1)\times SO(2,1)\rightarrow SO(2)$. 
\subsection{$SO(1,1)^{(2)}_\text{diag}\times SO(2,1)$}
We now consider another embedding for the $SO(1,1)$ factor as a diagonal subgroup of $SO(1,1)\times SO(1,1)$ generated by the non-compact generators $Y_{41}$ and $Y_{52}$ of $SO(5,5)$. The embedding tensor for $SO(2,1)$ remains the same as in the previous case. We will use the following non-vanishing components of the embedding tensor for the $SO(1,1)^{(2)}_{\textrm{diag}}$ 
\begin{equation}
 \xi_{46}=\xi_{57}=g_1\, .
\end{equation}
The scalar potential for this gauging is given by
\begin{eqnarray}
V = \frac{g_2^2 + g_1^2\Sigma^6}{2\Sigma^2}
\end{eqnarray}
with a unique critical point at
\begin{eqnarray}
\Sigma =\left(\frac{g_2}{\sqrt{2} g_1}\right)^{1/3}\, . 
\end{eqnarray}
We can again shift this critical point to the value $\Sigma=1$ by setting $g_2 = \sqrt{2}g_1$ and obtain the value of the cosmological constant
\begin{eqnarray}
V_0 = \frac{3}{2}g_1^2\, .
\end{eqnarray}
Scalar masses at this critical point are given by 
\begin{eqnarray}
\begin{array}{ccccccccc}
m^2L^2= & 16_{\times 2}, & 10_{\times 6}, & -8_{\times 1}, & 8_{\times 4}, & -6_{\times 2}, & 2_{\times 4}, & 0_{\times 8}\, .
\end{array}
\end{eqnarray}
Notice that all masses are quantized, and the $dS_5$ vacuum is unstable because of the negative mass values $-8$ and $-6$. The value $m^2L^2=16$ corresponds to that of $\Sigma$. Three of the eight massless scalars correspond to the Goldstone bosons of the symmetry breaking $SO(1,1)_\text{diag}\times SO(2,1)\rightarrow SO(2)$ at the vacuum. 
\subsection{$SO(1,1)^{(3)}_\text{diag}\times SO(2,1)$}
In this case, the embedding tensor for $SO(2,1)$ remains the same as the previous two cases while the $SO(1,1)$ is a diagonal subgroup of $SO(1,1)\times SO(1,1)\times SO(1,1)$. Non-vanishing components of the full embedding tensor are now given by
\begin{eqnarray}
\xi_{19} = \xi_{47} = \xi_{56}=g_1\qquad  \textrm{and}\qquad  f_{238} = -g_2\, .
\end{eqnarray}
These give the following scalar potential for $\Sigma$
\begin{eqnarray}
V = \frac{2g_2^2 + 3 g_1^2\Sigma^6}{4\Sigma^2}
\end{eqnarray}
which admits a critical point at
\begin{eqnarray}
\Sigma = \left(\frac{g_2}{\sqrt{3}g_1}\right)^{1/3}\, .
\end{eqnarray}
After setting $g_2 = \sqrt{3}g_1$, we find the cosmological constant
\begin{eqnarray}
V_0 =\frac{9}{4} g_1^2\, .
\end{eqnarray}
Scalar masses are given by
\begin{eqnarray}
\begin{array}{ccccccc}
m^2L^2=& 16_{\times 1}, & \dfrac{2}{3}(5+\sqrt{201})_{\times 3}, & \dfrac{2}{3}(5-\sqrt{201})_{\times 3}, & \left(\dfrac{28}{3}\right)_{\times 8},\\  &
8_{\times 2}, &\left(\dfrac{4}{3}\right)_{\times 3}, &0_{\times 8}   \, .
\end{array}
\end{eqnarray}
This $dS_5$ vacuum is unstable due to the negative mass value $\dfrac{2}{3}(5-\sqrt{201})$ with three massless scalars corresponding to the Goldstone bosons of the symmetry breaking $SO(1,1)_{\textrm{diag}}\times SO(2,1)\rightarrow SO(2)$.

\subsection{$SO(1,1)\times SO(2,1)\times SO(2,1)$}
In this case, the four noncompact generators of $SO(2,2)\sim SO(2,1)\times SO(2,1)$ group are embedded in the R-symmetry directions $M=1,2,3,4$ while the compact $SO(2)\times SO(2)$ generators are embedded in the matter directions $M=8,9$. The embedding tensor for the $SO(2,2)$ factor can be chosen as
\begin{eqnarray}
f_{238} = -g_2\qquad  \textrm{and}\qquad  f_{149} = -g_3\, .
\end{eqnarray}
The embedding tensor for $SO(1,1)$ is still given by \eqref{SO1_1_embedding}. Under the compact $SO(2)\times SO(2)$ symmetry, there are five singlet scalars corresponding to the non-compact generators $Y_{2a}$, $a=1,2,\ldots,5$. The coset representative can be written as
\begin{equation} 
\mc{V}=e^{\phi_1Y_{21}}e^{\phi_2Y_{22}}e^{\phi_3Y_{23}}e^{\phi_4Y_{24}}e^{\phi_5Y_{25}}\, .
\end{equation}
The scalar potential is given by
\begin{eqnarray}
V =\frac{1}{4\Sigma^2}\left[2(g_2^2+g_3^2)+ g_1^2\Sigma^6\cosh^2\phi_2\cosh^2\phi_3\cosh^2\phi_4\cosh^2\phi_5\right]
\end{eqnarray}
which admits only one critical point at
\begin{equation}
\phi_i=0,\quad i=1,2,\ldots, 5\qquad \textrm{and}\qquad \Sigma = \left(\frac{\sqrt{g_2^2+ g_3^2}}{g_1}\right)^{1/3}\, .
\end{equation}
The scalar mass spectrum is  
\begin{eqnarray}
m^2L^2=
\begin{array}{ccccccc}
&16_{\times 1} ,& 0_{\times 5}, & 4_{\times 12},& 2(3+\sqrt{17})_{\times 4}, & 2(3-\sqrt{17})_{\times 4}
\end{array}\, .
\end{eqnarray}
The $dS_5$ vacuum is unstable because of the negative mass value $2(3-\sqrt{17})$.
The mass value 16 corresponds to that of the dilaton while the five scalars with zero mass correspond to the Goldstone bosons of the symmetry breaking $SO(1,1)\times SO(2,1)\times SO(2,1)\rightarrow SO(2)\times SO(2)$. 

\subsection{$SO(1,1)\times SO(3,1)$}
The embedding tensor for the $SO(1,1)$ factor is the same as in \eqref{SO1_1_embedding} while the $SO(3,1)$ part is gauged by the following components of the embedding tensor
\begin{eqnarray}
f_{8,9,10}= f_{1,2,8}=f_{1,3,9}= f_{2,3,10}=-g_2 \, .
\end{eqnarray}
The compact $SO(3)\subset SO(3,1)$ is formed by generators $X_8$, $X_9$ and $X_{10}$ along the matter multiplet directions. 
\\
\indent There are five $SO(3)$ singlet scalars from $SO(5,5)/SO(5)\times SO(5)$ coset corresponding to the non-compact generators
\begin{eqnarray}
\tilde{Y}_1&=&Y_{15}-Y_{24}+Y_{33},\qquad \tilde{Y}_2=Y_{41},\nonumber \\
\tilde{Y}_3&=&Y_{42},\qquad \tilde{Y}_4=Y_{51},\qquad \tilde{Y}_5=Y_{52}\, .  
\end{eqnarray}
With the coset representative
\begin{equation}
\mc{V}=e^{\phi_1\tilde{Y}_1}e^{\phi_2\tilde{Y}_2}e^{\phi_3\tilde{Y}_3}e^{\phi_4\tilde{Y}_4}e^{\phi_5\tilde{Y}_5},
\end{equation}
the scalar potential is given by
\begin{eqnarray}
V&=&\frac{1}{128\Sigma^2}e^{-2(3\phi_1+\phi_2+\phi_3+\phi_4+\phi_5)}\left[8e^{2(\phi_2+\phi_3+\phi_4+\phi_5)}(1+3e^{4\phi_1}+16e^{6\phi_1}+3e^{8\phi_1}+e^{12\phi_1})\right.\nonumber \\
& &+4\sqrt{6}e^{3\phi_1+\phi_2+\phi_3+\phi_4+\phi_5}(e^{6\phi_1}+3e^{4\phi_1}-3e^{2\phi_1}-1)\left(e^{2\phi_2}-1+2e^{\phi_3+\phi_4}\right.\nonumber \\
& &-e^{2(\phi_3+\phi_4)}+e^{2(\phi_2+\phi_3+\phi_4)}+2e^{2\phi_2+\phi_3+\phi_4} -e^{2\phi_5}+e^{2(\phi_2+\phi_5)}-e^{2(\phi_3+\phi_4+\phi_5)}\nonumber \\
& &\left. +e^{2(\phi_2+\phi_3+\phi_4+\phi_5)}-2e^{\phi_3+\phi_4+2\phi_5}-2e^{2\phi_2+\phi_3+\phi_4+2\phi_5}\right)
\Sigma^3+3e^{6\phi_1}\left(1-2e^{2\phi_2}\right. \nonumber \\ 
& &+e^{4\phi_2}
-4e^{\phi_3+\phi_4}+6e^{2(\phi_3+\phi_4)}-4e^{3(\phi_3+\phi_4)}+e^{4(\phi_3+\phi_4)}
+4e^{2(\phi_2+\phi_3+\phi_4)} \nonumber \\
& &
+e^{4(\phi_2+\phi_3+\phi_4)}+6e^{2(2\phi_2+\phi_3+\phi_4)}+4e^{4\phi_2+\phi_3+\phi_4}-2e^{2(\phi_2+2\phi_3+2\phi_4)}
+2e^{2\phi_5}\nonumber \\
& &+4e^{4\phi_2+3\phi_3+3\phi_4}+e^{4\phi_5}-4e^{2(\phi_2+\phi_5)}+e^{4(\phi_2+\phi_5)}
-4e^{2(\phi_3+\phi_4+\phi_5)}+e^{4(\phi_3+\phi_4+\phi_5)}
\nonumber \\
& &-4e^{2(2\phi_2+\phi_3+\phi_4+\phi_5)}-4e^{2(\phi_2+2\phi_3+2\phi_4+\phi_5)}+2e^{4\phi_2+2\phi_5}+6e^{2(\phi_3+\phi_4+2\phi_5)}
\nonumber \\
& &+4e^{2(\phi_2+\phi_3+\phi_4+2\phi_5)}+6e^{2(2\phi_2+\phi_3+\phi_4+2\phi_5)}+2e^{4\phi_3+4\phi_4+2\phi_5}+2e^{4\phi_2+4\phi_3
+4\phi_4+2\phi_5}\nonumber \\
& &
+4e^{\phi_3+\phi_4+4\phi_5}+40e^{2(\phi_2+\phi_3+\phi_4+\phi_5)}-4e^{4\phi_2+\phi_3+\phi_4+4\phi_5}
-4e^{4\phi_2+\phi_3+\phi_4+4\phi_5}\nonumber \\
& &\left.\left.-2e^{2\phi_2+4\phi_5}+4e^{3\phi_3+3\phi_4+4\phi_5}
-4e^{4\phi_2+3\phi_3+3\phi_4+4\phi_5}
-2e^{2(\phi_2+2\phi_3+2\phi_4+2\phi_5)} \right) \right].\nonumber \\
& &
\end{eqnarray}
This potential admits a critical point at
\begin{eqnarray}
\Sigma = \left(\frac{\sqrt{3}g_2}{g_1}\right)^{1/3},\qquad \phi_i=0,\qquad i=1,2,\ldots,5, \qquad V_0 = \frac{3}{4}(3 g_1 g_2^2)^{2/3}\, .
\end{eqnarray}
We have not found other critical points from the above potential. As in other cases, we can bring this critical point to the value $\Sigma=1$ by setting $g_2 = \frac{g_1}{\sqrt{3}}$. 
\\
\indent The scalar mass spectrum at this $dS_5$ critical point is given by
\begin{eqnarray}
\begin{array}{ccccccccc}
 m^2L^2= &16_{\times 2}, & \dfrac{2}{3}(7+\sqrt{145})_{\times 3}, & \left[-\dfrac{20}{3}\right]_{\times 1}, & \dfrac{2}{3}(7-\sqrt{145})_{\times 3},\\ 
 & \left[\dfrac{16}{3}\right]_{\times 8}, & 4_{\times 4}, & 0_{\times 5}
\end{array}
\end{eqnarray}
which implies that the $dS_5$ vacuum is unstable because of the negative mass values $-20/3$ and $\frac{2}{3}(7-\sqrt{145})$. There are four Goldstone bosons corresponding to the symmetry breaking $SO(1,1)\times SO(3,1)\rightarrow SO(3)$.
\\
\indent We now move to a similar gauge group of the form $SO(1,1)^{(2)}_\text{diag}\times SO(3,1)$ with the embedding tensor 
\begin{equation}
\xi_{56}=\xi_{47} = g_1\qquad \textrm{and} \qquad f_{8,9,10}= f_{1,2,8}=f_{1,3,9}= f_{2,3,10}=-g_2\, . 
\end{equation}
Apart from a $dS_5$ critical point with all scalars from vector multiplets vahishing, we have not found any other critical point. Therefore, to simplify the expression, we will give the potential with only the dilaton $\Sigma$ non-vanishing   
\begin{equation}
V=\frac{1}{2\Sigma^2}(3g_2 ^2+ g_1^2 \Sigma^6).
\end{equation}
Choosing $g_2 = \sqrt{\frac{2}{3}}g_1$, the $dS_5$ critical point is given by
\begin{equation}
\Sigma = 1,\qquad V_0 = \frac{3}{2}g_1^2
\end{equation}
with the scalar mass spectrum
\begin{eqnarray}
\begin{array}{ccccccccc}
m^2L^2=& 16_{\times 1}, & \dfrac{4}{3}(5+\sqrt{73})_{\times 1}, & 10_{\times 6}, & \left(\dfrac{16}{3}\right)_{\times 5}, & \dfrac{4}{3}(5-\sqrt{73})_{\times 1}, \\ & 
 \left(-\dfrac{2}{3}\right)_{\times 6}, & 0_{\times 6} \, .
\end{array}
\end{eqnarray}
This $dS_5$ vacuum is again unstable because of negative mass values $-2/3$ and $\dfrac{4}{3}(5-\sqrt{73})$. The mass value $m^2L^2=16$ corresponds to that of $\Sigma$. 

\subsection{$SO(1,1)\times SU(2,1)$ }
We now consider the last gauge group that can be embedded in $SO(5,5)$. Non-vanishing components of the embedding tensor for the non-abelian part $SU(2,1)$ are given by
\begin{eqnarray}
f_{129} &=& f_{138}=f_{147} =f_{248}= -f_{349} = -f_{237}=g_2, \nonumber\\
f_{789} &=& -2g_2, \qquad f_{3,4,10} =f_{1,2,10} = \sqrt{3}g_2\, .
\end{eqnarray}
Gauge generators $(X_7,X_8,X_9)$ and $X_{10}$ correspond to the compact $SU(2)$ and $U(1)$, respectively.
\\
\indent Among the $25$ scalars in $SO(5,5)/SO(5)\times SO(5)$, there are two $SU(2)$ singlets corresponding to the following non-compact generators $Y_{51}$ and $Y_{55}$. The coset representative can be written as
\begin{equation}
\mc{V}=e^{\phi_1Y_{15}}e^{\phi_2Y_{55}},
\end{equation}
and, with the embedding tensor of $SO(1,1)$ given by \eqref{SO1_1_embedding}, the scalar potential reads
\begin{equation}
V =\frac{1}{16\Sigma^2}e^{-2\phi_2}\left[96g_2^2e^{2\phi_2}+(1+e^{2\phi_2})^2g_1^2\Sigma^6\right].
\end{equation}
There is only one critical point at
\begin{eqnarray}
\phi_2=0,\qquad \Sigma = \left(\frac{2\sqrt{3}g_2}{g_1}\right)^{1/3}, \qquad V_0 = 3 \left(\frac{3}{2}g_1 g_2^2\right)^{2/3}\, .
\end{eqnarray}
The scalar mass spectrum for all the 26 scalars is identical to that of $SO(1,1)\times SO(2,2)$ gauge group with the five massless scalars being the Goldstone bosons of the symmetry breaking $SO(1,1)\times SU(2,1)\rightarrow SU(2)\times U(1)$.

\subsection{$SO(1,1)\times SO(4,1)$}
As a final example, we consider $N=4$ gauged supergravity coupled to seven vector multiplets with $SO(1,1)\times SO(4,1)$ gauge group. The compact part of $SO(4,1)$ containing six generators of $SO(4)$ is embedded in the matter multiplet directions $M=7,...,12$ while the $4$ noncompact generators are embedded in the R-symmetry directions $M=1,2,3,4$. The embedding tensor for $SO(4,1)$ is then given by
\begin{eqnarray}
f_{127} =f_{138}=f_{1,4,10}=f_{239} = f_{2,4,11}=f_{3,4,12}=g_2,\nonumber\\
f_{789}=f_{7,10,11}=f_{8,10,12}=f_{9,11,12}=-g_2\, .
\end{eqnarray}
The embedding of $SO(1,1)$ is again given in \eqref{SO1_1_embedding}. 
\\
\indent Under $SO(4)\subset SO(4,1)$ symmetry, there are three singlet scalars from $SO(5,7)/SO(5)\times SO(7)$ coset corresponding to non-compact generators $Y_{51}$, $Y_{61}$ and $Y_{62}-Y_{71}$. Using the coset representative
\begin{equation}
\mc{V}=e^{\phi_1Y_{51}}e^{\phi_2Y_{61}}e^{\phi_3(Y_{62}-Y_{71})},
\end{equation} 
we find the scalar potential for this gauge group
\begin{eqnarray}
V &=& \frac{1}{64\Sigma^2}e^{-8\phi_1}\left[192g_2^2e^{8\phi_1}+g_1^2e^{4\phi_2}(1+2e^{4\phi_1}+e^{8\phi_1}-e^{4\phi_2}+2e^{4(\phi_1+\phi_2)}\right. \nonumber \\
& &\left.-e^{4(2\phi_1+\phi_2)})^2\Sigma^6\right]
\end{eqnarray}
which admits only one critical point at
\begin{eqnarray}
\phi_i=0,i=1,2,3,\qquad \Sigma= \left(\frac{\sqrt{6}g_2}{g_1}\right)^{1/3}, \qquad V =\frac{3}{2}\left(\frac{3 g_1 g_2^2}{\sqrt{2}}\right)^{2/3}\, .
\end{eqnarray}
Scalar masses at this vacuum are given by 
\begin{eqnarray}
\begin{array}{ccccccc}
m^2L^2= &16_{\times 1} ,& 0_{\times 5}, & 4_{\times 22},& 2(3+\sqrt{17})_{\times 4}, & 2(3-\sqrt{17})_{\times 4}&
\end{array}
\end{eqnarray}
which implies the instability of the $dS_5$ critical point because of the negative mass value $3-\sqrt{17}$. The five massless scalars are Goldstone bosons of the symmetry breaking $SO(1,1)\times SO(4,1)\rightarrow SO(4)$.
\section{Conclusions}\label{conclusions}
In this paper, we have studied $dS_5$ vacua of five-dimensional $N=4$ gauged supergravity coupled to vector multiplets. By using a simple ansatz for solving the extremization and positivity of the scalar potential, we have derived a set of general conditions for determining the form of gauge groups admitting $dS_5$ vacua as maximally symmetric backgrounds. In particular, these gauge groups must be of the form $SO(1,1)\times SO(2,1)\times G_{\textrm{nc}}\times H_{\textrm{c}}$ with $SO(1,1)$ gauged by one of the graviphotons that is $SO(5)_R$ singlet. $G_{\textrm{nc}}$ is a non-compact group whose compact subgroup is gauged by vector fields in vector multiplets. On the other hand, the compact and non-compact parts of $SO(2,1)$ are embedded along the R-symmetry and matter directions, respectively. $H_{\textrm{c}}\subset SO(n)$ is a compact group gauged by vector fields from the vector multiplets. We also note that the $SO(2,1)$ factor is not needed for the $dS_5$ vacua to exist.
\\
\indent The results provide a new approach for finding $dS_5$ vacua which could be of particular interest in various contexts such as in the dS/CFT correspondence and cosmology. We have also explicitly computed scalar potentials and studied $dS_5$ vacua for a number of gauge groups. However, as in the $N=4$ gauged supergravity in four dimensions, all of these $dS_5$ vacua are unstable. Although the conditions we have imposed are very restrictive, the analysis given here might provide a starting point for a systematic classification of $dS_5$ vacua in $N=4$ gauged supergravity. 
\\
\indent It would be interesting to carry out a similar analysis in gauged supergravities in other dimensions in particular in four-dimensional $N=4$ gauged supergravity and compare with the results given in \cite{de-Roo-Panda-1} and \cite{de-Roo-Panda-2}. Another direction would be to relate the known $dS_5$ vacuum of $N=8$ gauged supergravity with $SO(3,3)$ gauge group to $dS_5$ solutions of $N=4$ gauged supergravity considered here. On the other hand, embedding the $dS_5$ vacua identified here in $N=2$ gauged supergravity by truncating out some scalar fields might give stable $dS_5$ vacua within $N=2$ gauged supergravity as in a similar study in four-dimensional gauged supergravity \cite{Roest-09}. Finally, relations between $dS_4$ and $dS_5$ vacua in $N=4$ gauged supergravity similar to the $N=2$ case studied in \cite{ogetbil-2} also deserve further study. We hope to come back to these issues in future works.    
\vspace{0.5cm}\\
{\large{\textbf{Acknowledgement}}} \\
P. K. is supported by The Thailand Research Fund (TRF) under grant RSA6280022.

\appendix
\section{Useful formulae}
In this section, we collect some useful identities involving $SO(5)$ gamma matrices which are useful in the analysis of the constraints on the embedding tensor. For convenience, we also give a summary of non-vanishing components of the embedding tensor for gauge groups considered in the main text. 

\subsection{$SO(5)$ gamma matrices}
The $SO(5)$ gamma matrices ${\Gamma_{mi}}^j$ satisfy the Clifford algebra
\begin{equation}
{\Gamma_{mi}}^k{\Gamma_{nk}}^j+{\Gamma_{ni}}^k{\Gamma_{mk}}^j=2\delta^j_i\delta_{mn}\, .
\end{equation}
The charge conjugation matrix $C_{ij}$ and its inverse $C^{ij}$ are related to the $USp(4)$ symplectic form as follow
\begin{equation}
C_{ij}=-\Omega_{ij}\qquad \textrm{and}\qquad C^{ij}=\Omega^{ij}\, .
\end{equation}
We can now define the gamma matrices with all indices up or down
\begin{equation}
\Gamma_{m}^{ij}=\Omega^{ik}{\Gamma_{mk}}^j=-\Gamma^{ji}_m\qquad \textrm{and}\qquad \Gamma_{mij}={\Gamma_{mi}}^k\Omega_{kj}=-\Gamma_{mji}=(\Gamma_m^{ij})^*\, .
\end{equation}
Anti-symmetrized products of gamma matrices are defined as
\begin{equation}
\Gamma_{m_1m_2\ldots m_n}=\Gamma_{[m_1}\Gamma_{m_2}\ldots \Gamma_{m_n]}\, .
\end{equation}
In particular, we have the following relations
\begin{equation}
\Gamma_{mn}=\frac{1}{2}[\Gamma_m,\Gamma_n]\qquad\textrm{and}\qquad \Gamma_{mnp}=\frac{1}{2}\{\Gamma_m,\Gamma_{np}\}
\end{equation}
with the symmetry
\begin{equation}
(\Gamma_{mn})^{ij}=(\Gamma_{mn})^{ji}\qquad\textrm{and}\qquad (\Gamma_{mnp})^{ij}=(\Gamma_{mnp})^{ji}\, .
\end{equation}
Other useful identities are 
\begin{eqnarray}
\Gamma^{ij}_m \Gamma_{n ij} &=& 4\delta_{mn} \label{eq:a1},\\
\textrm{Tr}(\Gamma_{mn}\Gamma_{pq}) &=& 4(\delta_{mq}\delta_{np} - \delta_{mp}\delta_{nq}) ,\label{eq:a2}\\
\Gamma_{m} &=&\pm \frac{1}{24}\epsilon_{mnpqr}\Gamma^{npqr} \label{eq:a3},\\
\Gamma_{mn} &=&\mp \frac{1}{6}\epsilon_{mnpqr}\Gamma^{pqr} \label{eq:a4},\\
\Gamma_{mnp} &=&\mp \frac{1}{2}\epsilon_{mnpqr}\Gamma^{qr} \label{eq:a5},\\
\Gamma_{mnpq} &=&\pm\epsilon_{mnpqr}\Gamma^{r}\label{eq:a6},\\
\{\Gamma_{mn}, \Gamma_{pq}\}&=& 2\Gamma_{mnpq}+ 2\delta_{np}\delta_{mq} -2\delta_{nq}\delta_{mp},\label{eq:a7}
\end{eqnarray}
with $\epsilon_{12345}=+1$. The sign choices are related to the definition $\Gamma_5=\pm \Gamma_1\Gamma_2\Gamma_3\Gamma_4$. In our calculations, we choose the convention with the upper sign.
 
\subsection{Embedding tensor for some gauge groups with $dS_5$ vacua}
Examples of gauge groups that lead to $dS_5$ vacua are listed in table \ref{table:dS5} along with non-vanishing components of the corresponding embedding tensor. 

\begin{table}[!htbp]
\centering
\begin{tabular}{|c|c|c|}
\hline
 Gauge group & Embedding tensor &  Unbroken symmetry\\
\hline
$SO(1,1)\times SO(2,1)$ & $\xi_{56}=g_1,\quad f_{238} = -g_2$ &  $SO(2)$\\\hline
$SO(1,1)^{(2)}_\text{diag}\times SO(2,1)$ & $\begin{array}{l}\xi_{46}=\xi_{57} =g_1, \\f_{238} = -g_2\end{array}$ &  $SO(2)$\\\hline
$SO(1,1)^{(3)}_\text{diag}\times SO(2,1)$ & $\begin{array}{l}\xi_{19}=\xi_{56}=\xi_{47} =g_1, \\f_{238} = -g_2\end{array}$ & $SO(2)$\\\hline
 $SO(1,1)\times SO(2,1)^2$ & $\begin{array}{l}\xi_{56}=g_1, \\f_{238} =-g_2,\, f_{149} = -g_3\end{array}$  & $SO(2)\times SO(2)$\\\hline
$SO(1,1)\times SO(3,1)$ &$\begin{array}{l}\xi_{56}=g_1, \\f_{8,9,10}=-g_2, \\f_{128} = f_{139} = f_{2,3,10}=g_2\end{array}$ &$SO(3)$ \\\hline
$SO(1,1)^{(2)}_\text{diag}\times SO(3,1)$  &$\begin{array}{l}\xi_{46}=\xi_{57}=g_1, \\f_{8,9,10}=-g_2, \\f_{128} =f_{139} = g_2,\\ f_{2,3,10}=g_2\end{array}$ & $SO(3)$\\\hline 
 $SO(1,1)\times SU(2,1)$ & $\begin{array}{l}\xi_{56} =1,\\f_{129}=f_{138} =g_2,\\ f_{147} = f_{248} = g_2,\\ f_{349} = f_{237} = -g_2,\\ f_{789} = -2g_2, \\f_{3,4,10} = f_{1,2,10} = \sqrt{3}g_2\end{array}$  & $SU(2)\times U(1)$ \\\hline
$SO(1,1)\times SO(4,1)$ & $\begin{array}{l}\xi_{56} =1, \\f_{127} = f_{138} = f_{1,4,10} = g_2, \\ f_{239} =f_{2,4,11} = f_{3,4,12} = g_2,\\ f_{789} = f_{7,10,11}=-g_2,\\  f_{8,10,12} = f_{9,11,12} = -g_2,\end{array}$  &$SO(4)$ \\
\hline
\end{tabular}
\caption{Examples of gauge groups that give rise to $dS_5$ vacua in matter-coupled $N=4$ five-dimensional gauged supergravity are listed along with the corresponding embedding tensors. All of these gauge groups can be embedded in $SO(5,n)$ with $n\geq 7$.}\label{table:dS5}
\end{table}
\newpage 


\end{document}